\documentclass[twocolumn,showpacs,preprintnumbers,amsmath,amssymb]{revtex4}
\usepackage{graphicx}
\usepackage{dcolumn}
\usepackage{bm}

\begin{document}


\title{Gray soliton solution in the extended nonlinear Schr\"{o}dinger equation}

\author{ M~A~Borich }
\author{V~V~Smagin}
\author{A~P~Tankeyev}
 \affiliation{Institute of Metal Physics
of the Ural Division of the Russian Academy of Sciences, 18, Sofia
Kovalevskaya St. GSP-170 Ekaterinburg 620219, Russia}

\date{\today}

\begin{abstract}
In the framework of extended nonlinear Schr\"{o}dinger equation
(ENSE) the classification of self-similar solutions by the
relation between the amplitude and phase is performed. New
solutions of ENSE - ``gray soliton'' and ``gray soliton chain''
are presented. The properties of these solutions and the
possibility for using theirs in physical applications are
discussed.
\end{abstract}

\maketitle

The evolution of the waves in a weakly nonlinear higher order
dispersion media can be described by means of the extended
nonlinear Schr\"{o}dinger equation (ENSE)
  \begin{eqnarray}\label{ENLS}
  i\frac{\partial\varphi}{\partial t} &+&
\alpha\frac{\partial^2\varphi}{\partial
  y^2}+N |\varphi|^2\varphi + %
  \\\nonumber%
  &+& i\alpha_3\frac{\partial^3\varphi}{\partial
  y^3}  + i\alpha_1|\varphi|^2\frac{\partial\varphi}{\partial
  y} + i\alpha_2\varphi\frac{\partial}{\partial
  y}(|\varphi|^2)= 0.
  \end{eqnarray}
The first three terms in (\ref{ENLS}) form ``classic'' nonlinear
Schr\"{o}dinger equation (NSE). The coefficients of equation are
connected with properties of propagating waves. So,
$\alpha\sim\omega''_{kk}$ is the group velocity dispersion
($\omega$ is frequency and $k$ is wave number of carrier wave),
$\alpha_3$ describes the third-order dispersion, $N$ connects with
nonlinear response of the medium. The coefficients $\alpha_1$,
$\alpha_2$ describe the nonlinear dispersion characteristics of
medium and their nature depends on the problem under
consideration. For example, the appearance of such terms in
nonlinear optics is caused on inhomogeneous raman scattering (see,
for example \cite{Radhakrishnan1996}), in the nonlinear
ferromagnetodynamics the values $\alpha_1 = \alpha_2$
\cite{Borich2003} show the dependence of nonlinear response of
medium from wave number.

The Eq. (\ref{ENLS}) is not completely integrable in terms of
inverse scattering problem, but there are some integrable cases.
The simplest of them $\alpha_1 = \alpha_2 = \alpha_3 = 0$, when
equation (\ref{ENLS}) reduce to NSE. The another one is $\alpha_1
= \alpha_2 = 6\alpha_3$, in this case (\ref{ENLS}) reduces to
modified Korteweg - de Vries equation by means of Galileo
transformation. For $\alpha_2 = 0$ and $\alpha\alpha_1 =
3\alpha_3N$ (Hirota conditions), the ENSE reduces to the Hirota
equation \cite{Hirota1973} and for $\alpha = 1$, $N = 2$,
$2\alpha_1 = 3\alpha_3$, $4\alpha_1 = 3\alpha_3$ eq. (\ref{ENLS})
reduces to the Sasa - Satsuma \cite{Sasa1991} equation.

In general case (with arbitrary values of parameters) equation
(\ref{ENLS}) is not completely integrable, but some solutions (and
even soliton-like) are known. In our opinion, the most important
of them are the follows: ``light'' and ``dark'' Potasek - Tabor
solitons \cite{Potasek1991}, \cite{Grudinin1990}, so-called
``embedded'' and ``radiating'' solitons \cite{Karpman2001},
cnoidal states \cite{Smagin2004} and some specific solutions
(including ``soliton with tail'' and ``algebraic soliton'') which
were found in \cite{Gromov1998} (but these solutions realize only
for specific values of parameters $\alpha_1$, $\alpha_2$).

In this paper we try to extend the class of exact solution of ENSE
in non integrable case (with arbitrary values of coefficients). We
show here that a wide class of solutions can be classified in form
of the relation between the nonlinear frequency shift and
amplitude of solution.

We suppose solution of (\ref{ENLS}) in the form of a stationary
wave with phase modulation
\begin{eqnarray}\label{FormSolution}
\varphi(y, t) = F(y-vt)\exp\{i(p y - q t + \sigma(y-v t))\}.
\end{eqnarray}
In this case the equation (\ref{ENLS}) reduces to the system of
two ordinary differential equations (ODEs) for real functions
$F(x)$, $\Omega(x)$ (we denote here $x = y - v t$, $\Omega(x) =
\sigma'_x$)
\begin{eqnarray}%
\label{RealPartBaseEqn}%
& &(\alpha-3\alpha_3(\Omega + p)) \frac{d^2F}{dx^2} - 3\alpha_3
\frac{d F}{dx}\frac{d\Omega}{dx}+ %
\\\nonumber%
& &+[N - \alpha_1(p + \Omega)] F^3 + %
\\\nonumber%
& & + [-\alpha(\Omega + p)^2 + \alpha_3(\Omega + p)^3 -
\alpha_3\frac{d^2\Omega}{dx^2} + \Omega v + q]F = 0 %
\\\label{ImagPartBaseEqn}%
& &\alpha_3 \frac{d^3 F}{dx^3} + [2\alpha(\Omega + p) -
3\alpha_3(\Omega + p)^2 - v]\frac{d F}{dx} + %
\\\nonumber%
& &\qquad(\alpha_1 + 2\alpha_2)F^2\frac{d F}{dx} + (\alpha -
3\alpha_3(p + \Omega))F\frac{d\Omega}{dx} = 0.
\end{eqnarray}

Let us consider the solution of the system of equations
(\ref{RealPartBaseEqn}), (\ref{ImagPartBaseEqn}) in the following
form
\begin{eqnarray}%
\label{PhaseAmplRelation}%
\Omega = A + B F^n.
\end{eqnarray}
Note, that we do not impose restrictions on possible values of $n$
- it may be any real number, both positive and negative, integer
and fractional. Note also, that case $n = 1$ was discussed in
\cite{Gromov1998} and the case $B=0$ was considered in
\cite{Smagin2004}. By substituting (\ref{PhaseAmplRelation}) into
the system of ODEs (\ref{RealPartBaseEqn}),
(\ref{ImagPartBaseEqn}) and integrating the equation
(\ref{ImagPartBaseEqn}) with respect to $x$ we come to the system
\begin{widetext}
\begin{eqnarray}%
\label{RealPartNDegreeEqn}%
0&=& (\alpha - 3\alpha_3(A + p))\frac{d^2 F}{dx^2} + [N -
\alpha_1(A+p)]F^3 +[-\alpha(A + p)^2 + \alpha_3(A +
p)^3 + v A + q]F - %
\\\nonumber%
&-&(n+3)\alpha_3BF^n\frac{d^2 F}{dx^2} -
n(n+2)\alpha_3BF^{n-1}\Bigg(\frac{d F}{dx}\Bigg)^2 + [-2\alpha(A +
p) + 3\alpha_3(A + p)^2 + v]BF^{1+n} - \alpha_1BF^{3+n} + %
\\\nonumber%
&+&(-\alpha+3\alpha_3(A+p))B^2F(x)^{1+2n} + \alpha_3B^3F(x)^{1+3n},%
\\\label{ImagPartNDegreeEqn}%
0&=&\alpha_3\frac{d^2 F}{dx^2} + [2\alpha(A + p) - 3\alpha_3(A +
p)^2 - v]F + \frac{1}{3}(\alpha_1 +
2\alpha_2)F^3 + \frac{n+2}{n+1}(\alpha -3\alpha_3(A+p))B F^{1+n} - %
\\\nonumber%
&-& 3\frac{1+n}{1+2n}\alpha_3B^2F^{1+2n} + C_1.
\end{eqnarray}
\end{widetext}
Let's compare the Eq. (\ref{RealPartNDegreeEqn}) and Eq.
(\ref{ImagPartNDegreeEqn}). The first of them has two ``bad''
terms:
\begin{eqnarray}\label{FirstBadTerm}%
-(n+3)\alpha_3BF^nF''_{xx},%
\end{eqnarray}
which is proportional to the product of $n$ power of function $F$
by its second derivation, and
\begin{eqnarray}\label{SecondBadTerm}%
-n(n+2)\alpha_3BF^{n-1}(F'_x)^2,
\end{eqnarray}
 containing the product of $n-1$ power of $F$ and square of its first
 derivation. We can easy  get rid these terms in the following
 way. At first we multiply the equation (\ref{ImagPartNDegreeEqn}) by
 $(n+3)BF^n$ and add the result to the
 (\ref{RealPartNDegreeEqn}). Second, we multiply Eq.
 (\ref{ImagPartNDegreeEqn}) by $F'_x$, integrate the result,
 next find the relation for $(F'_x)^2$ from the obtained equation
 and substitute the result into  (\ref{RealPartNDegreeEqn}).
 Thereafter we have the system of two ODEs of the same form:
\begin{widetext}
\begin{eqnarray}\nonumber
0&=&(\alpha-3\alpha_3(A+p))\frac{d^2 F}{dx^2} + [-\alpha(A+p)^2 +
\alpha_3(A + p)^3+ v A + q]F + [N -\alpha_1
(A + p)]F^3 + 2n(n+2)C_2BF^{n-1} + %
\\\nonumber %
&+&(n^2+3n+2)[2\alpha(A+p) - 3\alpha_3(A+p)^2-v]BF^{1+n}+
(3n+5)(\alpha - 3\alpha_3(A+p))B^2F^{1+2n} + %
\\\label{FirstEqn}%
&+&(2n^2+5n+3)BC_1F^n+[(\alpha_1 +2\alpha_2)\frac{n^2}{6}
+(\alpha_1+2\alpha_2)\frac{2n}{3}+2\alpha_2]B F^{3+n} -
2\frac{3n^2 + 8n + 4}{1+2n}\alpha_3B^3F^{1+3n}
\\\nonumber
0&=&\alpha_3\frac{d^2 F}{dx^2} + [2\alpha(A + p) - 3\alpha_3(A +
p)^2 - v]F + \frac{1}{3}(\alpha_1 + 2\alpha_2)F^3 + \frac{n+2}{n+1}(\alpha -3\alpha_3(A+p))B F^{1+n} +%
\\\label{SecondEqn}%
&-& 3\frac{1+n}{1+2n}\alpha_3B^2F^{1+2n} + C_1,
\end{eqnarray}
\end{widetext}
where $C_1$ and $C_2$ are constants of integration. The system of
this ODEs has nontrivial solutions if the equations
(\ref{FirstEqn}), (\ref{SecondEqn}) coincide, that is they have to
contain the value $F(x)$ with same indices of power and
proportional coefficients. In this context there are two cases
available: (i) all exponents in (\ref{FirstEqn}),
(\ref{SecondEqn}) are different and (ii) some values coincide.

(i). If all exponents have different values, the conditions of
compatibility of the system (\ref{FirstEqn}), (\ref{SecondEqn})
have the follows form:
\begin{eqnarray}\label{CompCondGenN}
F^n, F^0&:&C_1=0%
\\\nonumber%
F^{n-1}&:&C_2=0%
\\\nonumber%
F^{3+n}&:&B\Big[\frac{n^2}{6}(\alpha_1 + 2\alpha_2) +
\frac{2n}{3}(\alpha_1 + 2\alpha_2) + 2\alpha_2\Big] = 0%
\\\nonumber%
F^{1+3n}&:&\alpha_3B^2[3n^2+8n+4] = 0.
\end{eqnarray}
We don't consider here the case $B = 0$ - in this case we
immediately come to the Duffing equation and brief investigation
of available solutions in this situation can be found in
\cite{Smagin2004}. It is clear from last two equations in
(\ref{CompCondGenN}) that system with $B\ne 0$ may have nontrivial
solutions in the only one case:
\begin{eqnarray}%
\label{CondNm23n}%
n &=& -2/3,%
\\\label{CondNm23alpha}%
\alpha_1 &=& 17/5\alpha_2.
\end{eqnarray}
So we have remarkable result: extended NSE has nontrivial
solutions only for bounded set of numbers $n$. The equation for $n
= -2/3$ has the follow form:
\begin{eqnarray}%
\label{EqNm23}%
0= \alpha_3\frac{d^2F}{dx^2} &+& \frac{9}{5}\alpha_2F^3 + 9(\alpha
- 3\alpha_3p)k_0F + %
\\\nonumber%
&+&4(\alpha - 3\alpha_3p)B F^{1/3} + 3\alpha_3B^2F^{-1/3},
\end{eqnarray}
with parameters
\begin{eqnarray}\label{EqNm23Pars}%
k_0 &=& \frac{15\alpha_3N - 17\alpha\alpha_2}{10\alpha_2\alpha_3},%
\\\nonumber%
A+p &=& -\frac{5\alpha_3N - 9\alpha\alpha_2}{10\alpha_2\alpha_3},%
\\\nonumber%
v &=& 2\alpha(A+p) - 3\alpha_3(A+p)^2 - 9(\alpha - 3\alpha_3p)k_0,%
\\\nonumber%
q &=& \alpha(A+p)^2 - \alpha_3(A+p)^3 - A v +%
\\\nonumber%
&+&k_0(2\alpha(A+p) - 3\alpha_3(A+p)^2-v).
\end{eqnarray}
At first sight the existence of physical system with singled out
parameters (\ref{CondNm23n}), (\ref{CondNm23alpha}) is low -
probability.

(ii). Let's investigate now the case, when there are some
coincident values among the indices of power of function $F(x)$ in
(\ref{FirstEqn}), (\ref{SecondEqn}). It appears when $n$ takes on
one of the values
\begin{eqnarray}\label{NValuesAvlb}%
n &=& 0,\quad 1, \quad 2, \quad 3,\quad 4,\quad -1, \quad -2,\quad
-3,\\\nonumber & &\quad -1/2,\quad -1/3,\quad 2/3.
\end{eqnarray}
For each value from (\ref{NValuesAvlb}) we have its own system of
equations and all cases should be analyzed separately. Note, the
case $n=0$ is identical to the case $B = 0$ (we can rename
$A+B\rightarrow A$ in such case). The values $n = -1, -1/2$ are
specific because of zeros in denominator. For this values we are
to do full transformation chain (\ref{RealPartBaseEqn}) -
(\ref{SecondEqn}). As for other values from (\ref{NValuesAvlb})
the final system of ODEs can be found after direct substitution
the corresponding value of $n$ into (\ref{FirstEqn}) and
(\ref{SecondEqn}).

After do this we come to the follow conclusion. Nontrivial
solutions may be found only in three cases:
\begin{enumerate}
    \item B = 0. Duffing equation. The classification of possible
    solutions was performed in \cite{Smagin2004}.
    \item n = 1. The brief classification of possible
    states in this situation was published in
    \cite{Gromov1998}.
    \item n = -2. This case will be discussed bellow.
\end{enumerate}
The substitution $n = 2,3,4,-1,-3,-1/2,-1/3,2/3$ requires the
condition $\alpha_3 = 0$. As result we have instead of third-order
differential equation a second-order ODE. We are not interesting
this situation in our paper.

\begin{figure}
  \includegraphics[width=7.0cm]{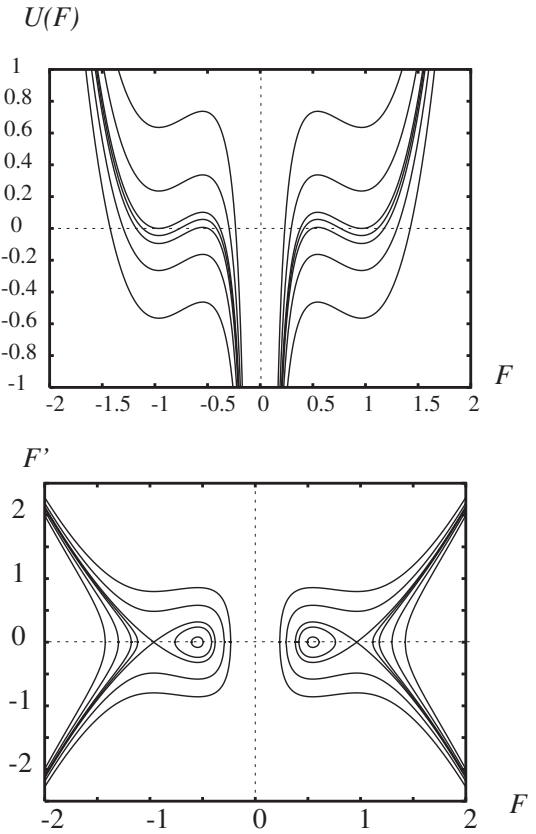}\\
  \caption{Phase portrait for the case $(\alpha_1 + 2\alpha_2)\alpha_3<0$}\label{FigPhasePort}
\end{figure}

In the case $n = -2$ we have the equation
\begin{eqnarray}\label{EqNm2}
 \frac{d^2F}{dx^2} + \alpha_3^{-1}(2\alpha(A+p) -
3\alpha_3(A+p)^2-v)F &+& %
\\\nonumber%
 + 1/3\alpha_3^{-1}(\alpha_1 + 2\alpha_2)F^3 - B^2F^{-3}
&=& 0
\end{eqnarray}
whose parameters satisfy the relations
\begin{eqnarray}\label{EqNm2Pars}
C_1 &=& 0, \\\nonumber%
&& \frac{\alpha - 3\alpha_3(A+p)}{\alpha_3} =
3\frac{N-\alpha_1(A+p)}{\alpha_1 + 2\alpha_2} =%
\\\nonumber%
&=&\frac{\frac{2}{3}(\alpha_2 - \alpha_1)B - \alpha(A+p)^2 +
\alpha_3(A+p)^3 + v A + q}{2\alpha(A+p) - 3\alpha_3(A+p)^2-v}.
\end{eqnarray}
The first integral of motion for this equation can be easy
calculated
\begin{eqnarray}\nonumber
 \Bigg(\frac{d F}{dx}\Bigg)^2 &=& -\alpha_3^{-1}(2\alpha(A+p) -
3\alpha_3(A+p)^2-v)F^2 - %
\\\label{EqNm2Int}%
&-& 1/4\alpha_3^{-1}(\alpha_1 + 2\alpha_2)F^4 - B^2F^{-2} + 2E.
\end{eqnarray}
Let's consider now the case $(\alpha_1 + 2\alpha_2)\alpha_3<0$. In
this situation potential (right - hand part of (\ref{EqNm2Int}))
is negative for small values of $F$ and increases infinitely for
large $F$. So, for small enough $F$ equation (\ref{EqNm2}) have no
real solution, whereas for large $F$ the solutions have infinite
trajectories. For some values of $E$ potential has three real
roots and the finite periodical solutions are possible. This
potential for different values of $E$ and corresponded phase
portrait is shown in fig. \ref{FigPhasePort}.
The solutions, corresponded to the finite trajectories in phase
portrait on fig. \ref{FigPhasePort} can be defined by means of
Jacobi elliptic delta function
\begin{eqnarray}\label{Nm2DNSol}
F(x) &=& F_0\Big\{1 - \frac{\mu^2}{1 + \mu^2}\mathrm{dn}(bx,
k)^2\Big\}^{1/2}
\end{eqnarray}
with parameters
\begin{eqnarray}\label{Nm2DNSolPars}
F_0&=&\Big\{-
\frac{2\eta(1+\mu^2)}{(3+\mu^2+k^2\mu^2)\xi}\Big\}^{1/2},%
\\\nonumber%
b&=& \mu\ \Big\{\frac{\eta}{3+\mu^2+k^2\mu^2} \Big\}^{1/2},%
\\\nonumber%
E &=& -\frac{\eta^2}{\xi}\frac{\mu^4k^2+2\mu^2+2k^2\mu^2+3}
{(3+\mu^2+k^2\mu^2)^2},%
\\\nonumber%
B^2 &=& \frac{4\eta^3}{\xi^2}\frac{\mu^4k^2+\mu^2+k^2\mu^2+1}
{(3+\mu^2+k^2\mu^2)^3},
\end{eqnarray}
where $\xi$, $\eta$ are coefficients standing before $F^3$ and $F$
in (\ref{EqNm2}) respectively. The separatrix solution (potential
has the point of contact with F axis, see fig. \ref{FigPhasePort})
appears when $k = 1$ ($0\le k\le 1$ - parameter of elliptic Jacobi
function). Returning to the space variable $y$ and to the function
$\varphi(y,t)$ and using the relations (\ref{Nm2DNSolPars}),
(\ref{EqNm2Pars}) we find gray soliton solution of extended NSE

\begin{figure}
  \center\includegraphics[width=7.0cm]{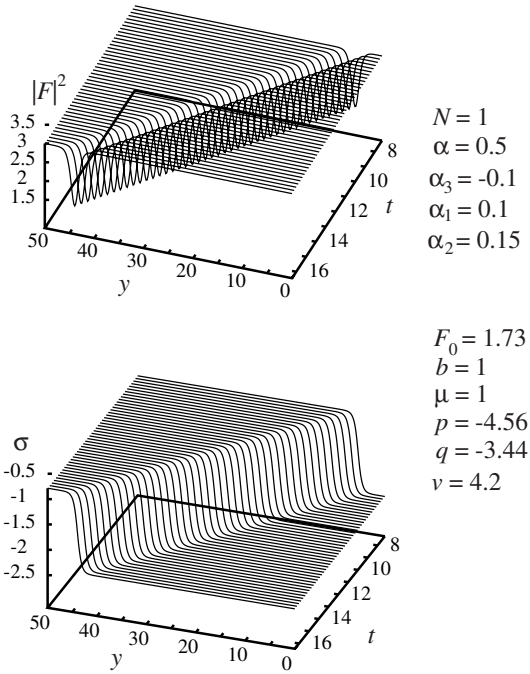}\\
  \caption{``Gray'' soliton in ENSE model} \label{FigGraySoliton}
\end{figure}

\begin{eqnarray}\label{Nm2GraySoliton}
\nonumber%
\varphi(y, t)&=&F(y-vt)\exp\{i(p y - qt + \sigma(y-vt))\},%
\\\nonumber%
F(y-vt) &=& F_0\Big\{1 -
\frac{\mu^2}{1+\mu^2}\mathrm{sech}^2(b(y-vt)) \Big\}^{1/2},%
\\\nonumber%
\sigma(y-vt) &=& \arctan[\mu\tanh(b(y-vt))],%
\\
F_0^2 &=& -\frac{6\alpha_3}{\alpha_1 +
2\alpha_2}\frac{1+\mu^2}{\mu^2}b^2,%
\\\nonumber%
p &=& \frac{\alpha(\alpha_1 + 2\alpha_2) -
3N\alpha_3}{6\alpha_3\alpha_2} + \frac{b}{\mu},%
\\\nonumber%
q &=& \alpha p^2 - \alpha_3p^3 - F_0^2(N-\alpha_1p),%
\\\nonumber%
v &=& 2\alpha p - 3\alpha_3p^2 - 2\alpha_3b^2 - 2(\alpha -
3\alpha_3p)b/\mu %
\\\nonumber%
&-& 6\alpha_3b^2/\mu^2.
\end{eqnarray}
The gray soliton solution of ``classic'' nonlinear Schr\"{o}dinger
equation was known long ago (see for instance \cite{Hasegawa1973})
but such solution in the ENSE was not known. The evolution of this
solution is shown in fig. \ref{FigGraySoliton}. For numerical
simulation we use Fourier transform in $y$-space and fourth-order
Runge-Kutta method in $t$- space.

In spite of Potasek - Tabor solitons \cite{Potasek1991} this
solution has two free parameters (for example thickness $b^{-1}$
and deep of modulation $\mu^2/(1+\mu^2)$, $0<\mu<\infty$). Note
also, that amplitude and phase in this solution are connected by
means the relation (\ref{PhaseAmplRelation}). So we can control
the deep of modulation by means of nonlinear phase shift, and
otherwise we can generate different phase shift using the
corresponded deep of modulation. These properties should be
fruitful for different physical application.

As for cnoidal solutions (\ref{Nm2DNSol}) at $k\rightarrow 1$ they
represent well separated holes on against a background of the
carrier wave and can be understand as ``1D gray soliton lattice''.
Here we discuss the case $(\alpha_1+2\alpha_2)\alpha_3<0$. In the
opposite case equation (\ref{EqNm2}) also has finite periodical
solutions but this situation should be discussed separately.

The authors wish to thank A. Shagalov for fruitful discussions.
The work was done within the framework of Program of Basic
Researches of the Presidium of RAS ``Mathematical method of
nonlinear dynamics'' and was partially supported by Grant of young
scientists and aspirants of Ural Division of RAS No M-06-02-05.

\end{document}